# ULTRA LOW-FREQUENCY OSCILLATIONS OF A SOLAR FILAMENT IN Hα REVEALED WITH THE DATA OF THE GLOBAL OSCILLATION NETWORK GROUP (GONG)

[1]Efremov V.I., [1]Parfinenko L.D., [1,2]Solov'ev A.A.

*Central (Pulkovo) Astronomical Observatory of RAS, St-Petersburg, Russia*

*Kalmyk State University, Elista, Russia*

**Abstract** *The data of ground-based telescopes of Global Oscillation Network Group (GONG) obtained in the H-alpha line provide an opportunity to study the long-period oscillations of chromospheric filaments (quiescent prominence). For the first time, on the base of time-series of 5 days duration, combined from the observations of three observatories of the GONG, a new ultra-low mode, with period between 20 and 30 hours, was reliably detected in oscillations of a long-lived dark filament on the solar disc.*

## 1. Introduction

The oscillations and wave propagations in solar prominences and coronal filaments are the focus of today's solar research. It suffices to mention the close connection of eruptive prominences with coronal mass ejections, which have a strong impact on the near-Earth space (Gopalswamy 2003; Foullon 2007 and many others). Solar prominences or filaments present cold and dense plasma structures built-in hot and rarefied corona. They consist of small-scale threadlike fibrils, the geometry, movement and thermal conductivity of which are controlled by the magnetic field. The filaments "hang" in the corona, supported against the gravity by magnetic fields anchored in the photosphere (Tandberg-Hanssen 1995). The filaments are usually located over the line of the polarity inversion of the magnetic field on the photosphere. The prominences are observed in emission on the solar limb, and in the absorption, on the disc, they appear as dark filaments.

The first systematic studies of oscillations in prominences have been carried out by Ramsey and Smith (1966). The oscillations of prominences are divided into two groups depending on the amplitude of the observed oscillations, namely, the vibrations with a large amplitude and small-amplitude oscillations (see e.g., Oliver 2002). The oscillations of large amplitudes are observed when prominences are experiencing large horizontal displacements from its equilibrium position (up to the $4 \times 10^4$ km), and prominence as a whole fluctuates with velocity amplitude of the order of 20 km/s or more. There are very few observations of prominences oscillations with a large displacements. Much more often the oscillations with small amplitude of the speed, 2-3 km/s or less, are observed (Oliver 2002). The periods of these oscillations lie between 1 and 80 minutes, and they are classified as follows: oscillations with a period of more than 40 minutes are called the long-period fluctuations; oscillations with periods between 40 and 10 minutes are called the average periodicals, and, finally, the vibrations of prominences with periods of less than 10 minutes are named the short-period oscillations. This classification is rather conditional, it has no physical basis (Oliver 2002). We intend in this paper to describe the oscillations of solar filaments with periods of 1-2 up to 20-30 hours. Such oscillations are revealed for the first time, their periods are far beyond the above classification. From the point of view of this classification, it is appropriate to call such phenomena as the 'ultra-long-period' or 'ultra-low-frequency' oscillations.

## 2. Data of observations

Oscillations of the filaments with periods of ten hours and more can not be detected by the normal ground-based observations because of the short duration of observations (the daylight). For the first time, it was reported about the ultra-long-periodic (8-27 hours) fluctuations in the EUV filaments by Foullon, Verwichte & Nakariakov (2004). They used the data from the space telescope SOHO/EIT in wavelength 195A. The consequent six daily series of observations were obtained with cadence of 12 minutes. The time-variations of the intensity of a coronal filament have been studied. The dominant period of 12.1 hours was derived for the oscillations with the intensity of the vibration amplitude of about 10% of the background.

Unfortunately, the space observations are not carried out in line Hα, wherein the chromospheric filaments can best be seen. However, the data in this line are available in the GONG - Global Oscillation Network Group (Hill 1994). This network consists of six identical terrestrial solar telescopes spaced longitudinally to provide continuous monitoring of the Sun around the clock. Scale of the image is 75 arc seconds per inch, diameter of lens is 7cm, CCD camera has 2048x2048 pixels. Images of the full disk of the Sun in the line Hα 656 nm are obtained by the GONG network at the middle of 2010 as an additional information to the data of basic system of 6 helioseismological telescopes.

The data of GONG network are repeatedly overlapped and have a local features, specific for each observatory. It is reflected in the gauge as for the magnetic field strength as for the intensity of Hα filtergrams. There are many options of overlappings, so we should make the choice of data to generate the time-series of intensity variations in the filament, for reasons of a minimum of observatories involved. In terms of computing process, it is the most suitable option. As a result, we have selected the full observations of two stations {Uh (The Udaipur Solar Observatory in India)}, {Bh (The Big Bear Solar Observatory in California, USA)} and a fragment of observations of the station {Th (The Observatorio del Teide in the Canary Islands)}. Figure 1 shows schematically (by color) the observations of selected observatories (blue), the absence of observations (red) and one possible cross-linking of the data (purple), in accordance with the above principle.

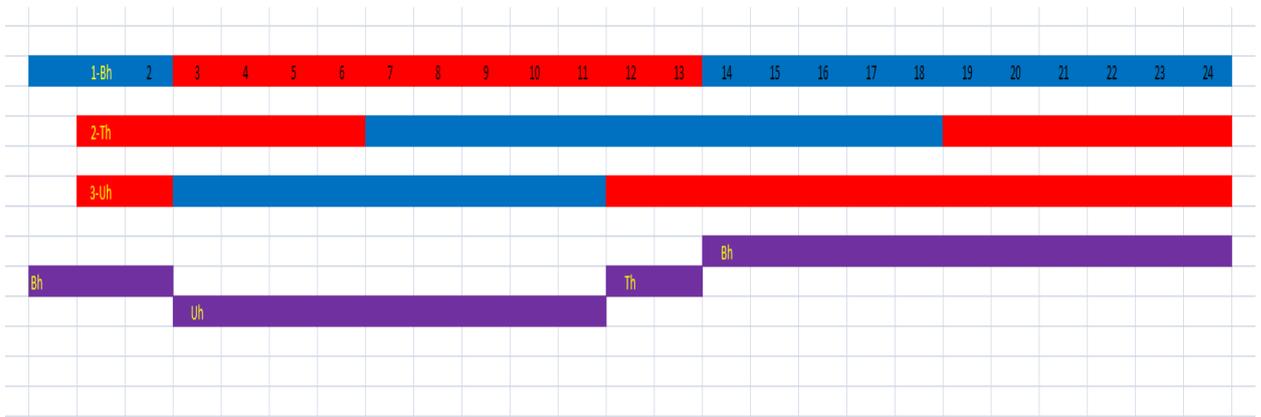

Fig.1. Pattern of choice of observatories and formation of corresponding time-series of the data.

## 3. RESULTS OF THE DATA PROCESSING

We processed the 120 hour time-series of Hα filtergrams, obtained by the telescopes of the GONG 2015/05/25-29. Figure 2 shows the very long filament, which, as it turns out on closer examination, is not a single entity; in fact, it consists of three parts, loosely connected or not connected with each other. As a result, the spectral characteristics in its three fragments were different. In the first Region Of Interest (ROI1), the obtained period of oscillations of filament intensity was 20 hours (Figure 3), for the points in the ROI 2 the period turned to be 22 hours, while for pixels in the portion 3 (ROI 3), the oscillation period was found to be 30 hours (Figure 4). Apparently, the difference is caused by the fact that the physical parameters (mass density and magnetic field) in the three parts of the extended filament differ markedly.

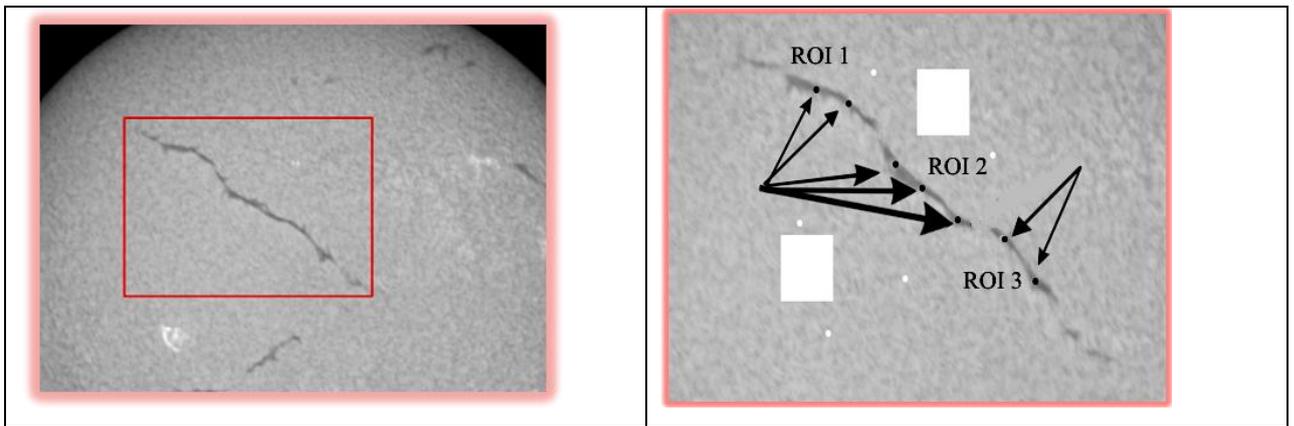

Fig.2. Left: Hα filtergram; the test fiber is allocated by red square. Right: the points at which the change of intensity are measured, into the filament (black dots) and outside it (white dots). In the white squares the average value of the intensity of the background (quiet photosphere) and its time-changes were studied.

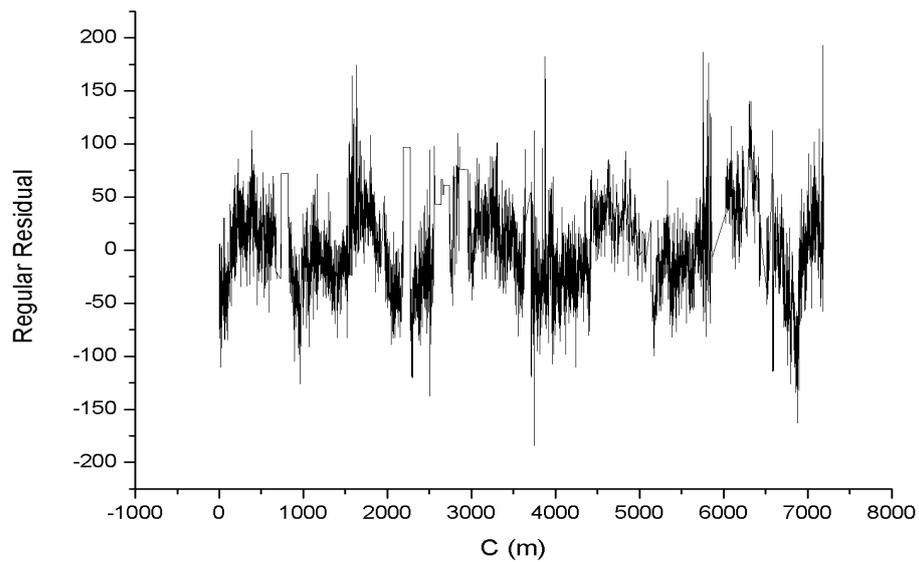

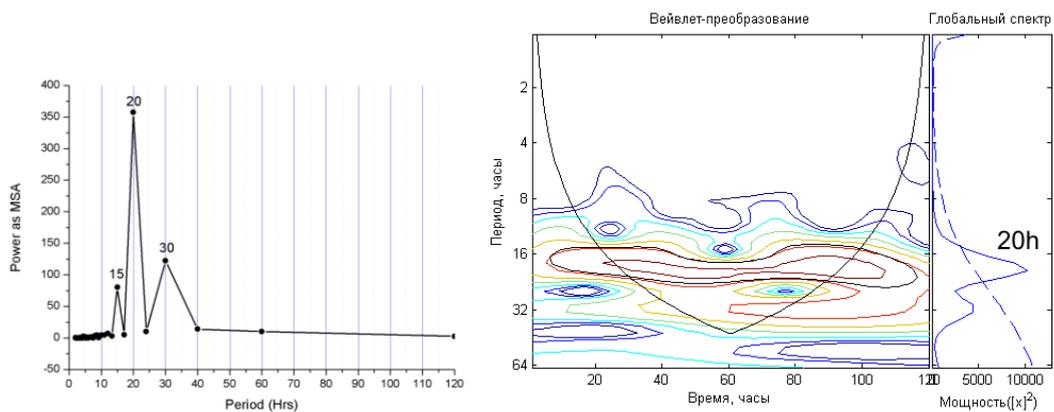

Fig.3. Top: the intensity at one of the points of the filament in ROI 1 area.     Bottom: the periodogram and Morlet wavelet. The period of oscillation is about 20 hours; as it can be best seen in wavelet, it is stable for all time of observation.

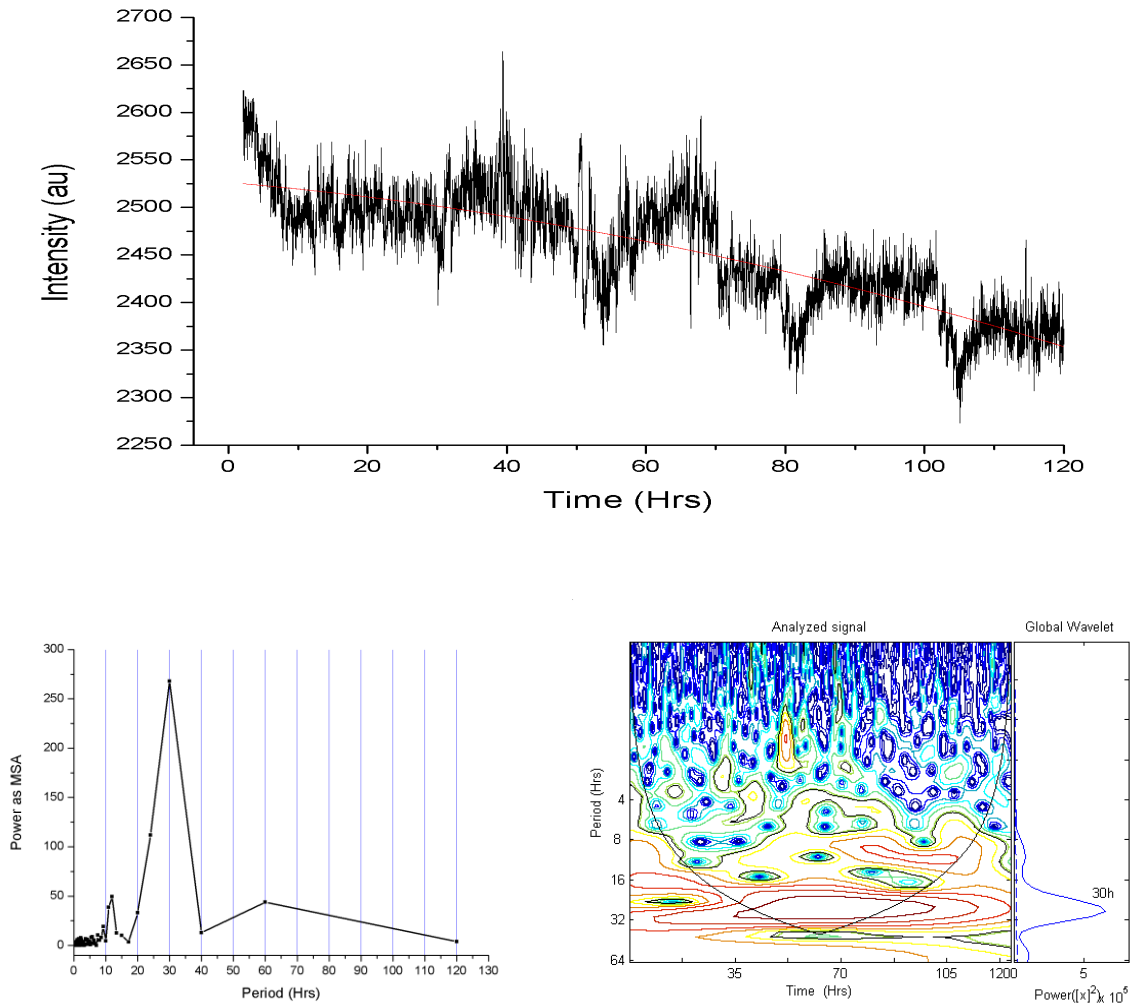

Fig.4. Top: the intensity at one of the points of the filament, ROI 3 area.  Bottom: the periodogram (FFT) and Morlet wavelet. The period of oscillation is about 30 hours; as it can be best seen in wavelet, it is stable for all time of observation.

No doubts, we were also interested in the processes occurring near the filament. There were some concerns related to the possible manifestation of the local features of observatories, and as a result, the appearance in the researched time-series the daily, 24-hour, artifact. We were also interested in how quickly the oscillating power falls with the distance from the filament. For this study we selected the number of test points: as well as close to the filament and substantially away from it. Also we have investigated the characteristics averaged over two white squares in vicinity of the filament (Fig. 2, right: white dots and squares). It was found that the oscillatory power decreases rather quickly with distance: near the filament it is an order less than the power of oscillations inside it. In Figure 5 we present, for comparing, the power spectrum of the oscillatory process in ROI 1 (red) and a number of spectra for the time series of intensity, built in for the near and far points outside the filament. Obviously, the power of oscillations decreases quickly with distance from the filament and has no significant component outside it. At the same time, the values of the periods of oscillations in ROI 1 (20h) and in ROI 3 (30h) differ greatly from the possible 24-hour artifact. Their reliability of 95% is marked, respectively, in the global wavelet spectrum (Figure 3, 4, bottom-right).

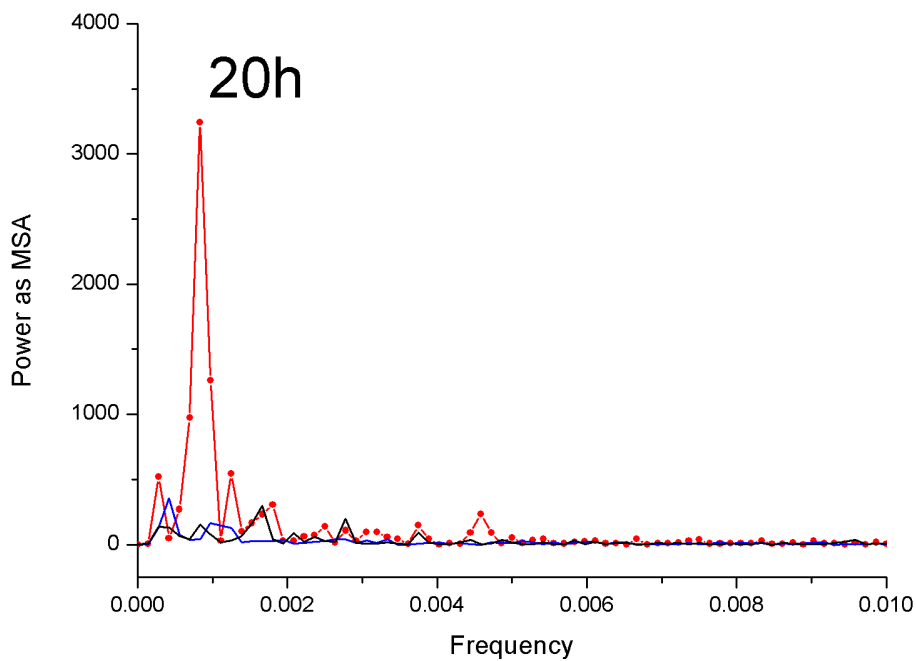

Fig.5. The power spectrum of the oscillatory process in ROI 1 (red) and a number of spectra for the series, built in for the near and far points outside the filament.

The resulting spectral composition of the filament oscillations has a very complex structure. In this spectrum, along with ultra-low mode, the mid-range mode (MM) with a typical average period of about 100-110 minutes drew the attention. This MM mode is important as it is well observed in the emission of radio sources associated with sunspots, which gives another confirmation of its solar origin. This problem is always encountered in the interpretation of ground-based observations. After the filtration of low frequency modes, let us consider one of the fragments of 700 minutes duration, randomly taken from the total time-series of the filament in the ROI 1. The result is shown in Fig. 3.

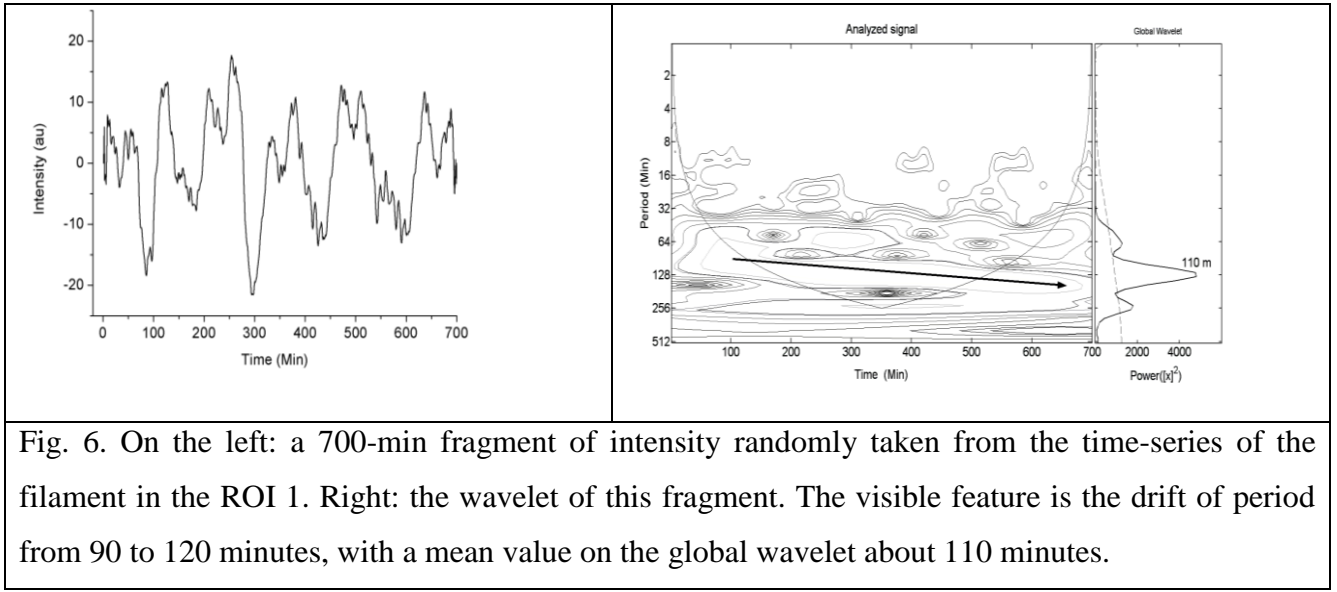

Fig. 6. On the left: a 700-min fragment of intensity randomly taken from the time-series of the filament in the ROI 1. Right: the wavelet of this fragment. The visible feature is the drift of period from 90 to 120 minutes, with a mean value on the global wavelet about 110 minutes.

### 4.Conclusions

1. For the first time, the continuous long-term ground-based observations of the GONG, produced in H-alfa line, allow to find the ultra low-frequency oscillations of the intensity of dark chromospheric filaments on the Sun with the periods in the range of 20-30 hours.

2. The mid-range mode (MM) with a period of 80-120 minutes, a well-known from observations of the radio sources associated with sunspots, was also revealed. This mode is persistent in the power spectrum of filament oscillation, but some slow drift of the period was detected. It can obviously be caused by relatively slow time-variations of the filament parameters, in particular, by the change of its magnetic field.

3. On the basis of our long-term studies of low frequency oscillations of solar magnetic structures (Efremov et al. 2010, 2014; Smirnova et al. 2013a,2013b; Bacunina et al. 2013), we conclude that a wide class of active elements on the Sun (sunspots, faculae, chromospheric filaments, coronal loops), being the long-lived objects, are in a state of stable magnetohydrostatic equilibrium, and can oscillate as a whole with respect to this state due to perturbations from outside. These eigen oscillations of a single magnetic object in the gravity field are, normally, rather slow. Large value of the periods of these oscillations is caused by the fact that the mass of the system is quite large, and the restoring force is not very strong, due to relatively low strength of its magnetic field (Solov'ev & Kirichek, 2014).

**Acknowledgements**

The authors thank the team of the GONG project for the opportunity to use their materials to solving our problems.